\documentclass[twocolumn,showpacs,prc]{revtex4}
\usepackage{graphicx}
\usepackage{dcolumn}
\usepackage{bm}
\begin{document}

\title{The Faddeev Model and Scaling in Quantum Chromodynamics}
\author{A. Widom and J. Swain}
\affiliation{Physics Department, Northeastern University, Boston MA USA}
\author{Y. N. Srivastava}
\affiliation{Physics Department, University of Perugia, Perugia Italy}

\begin{abstract}
The Faddeev two body bound state model is discussed as an example of a QCD 
inspired model thought by some to exhibit dimensional transmutation. This 
simple model is solved exactly and the growth of a specified dimensional 
energy scale is shown to be an illusion.  
\end{abstract}

\maketitle

\section{Introduction \label{intro}}

Consider a two spatial dimensional non-relativistic attractive 
potential scattering model\cite{Faddeev:2002,Jackiw:2000} 
described by the Faddeev Hamiltonian 
\begin{equation}
{\cal H}=-\left(\frac{\hbar^2}{2\mu }\right) 
\left[\Delta +\epsilon \delta ({\bf r})\right] 
\label{intro1}
\end{equation} 
wherein \begin{math} \mu \end{math} is the reduced mass of the two 
particles, \begin{math} \Delta \end{math} is the two dimensional 
Laplacean, \begin{math} \epsilon \end{math} is a dimensionless 
coupling strength and \begin{math} \delta ({\bf r}) \end{math} is 
a pure s-wave short ranged potential. In principle, one solves the 
Schr\"odinger equation for scattering states 
\begin{equation}
{\cal H}\psi ({\bf r})=E\psi ({\bf r})=
\left(\frac{\hbar^2k^2}{2\mu }\right)\psi ({\bf r})  
\label{intro2}
\end{equation} 
which behaves as 
\begin{equation}
\psi ({\bf r}) \to e^{ikx}+
\left(e^{ikr}\sqrt{\frac{i}{r}}\right)f(E) + \cdots  
\ \ \ {\rm as}\ \ \ r\to \infty .
\label{intro3}
\end{equation} 
The differential s-wave target length is thereby 
\begin{equation}
\left(\frac{dL}{d\theta }\right)=|f(E)|^2,
\label{intro4}
\end{equation} 
with a total target length given by the two dimenisonal 
optical theorem 
\begin{equation}
L=\sqrt{\frac{8\pi}{k}}\ {\Im m} f(E).
\label{intro5}
\end{equation} 
Introducing a complex energy in the upper half plane 
\begin{math} {\Im m}\ z>0  \end{math}, the dimensionless 
analytic scattering amplitude 
\begin{math} \tau (z) \end{math} determines 
\begin{math} f(E)  \end{math} via  
\begin{eqnarray}
f(E)=-\sqrt{\frac{1}{8\pi k}}\ \tau (E+i0^+), 
\nonumber \\ 
L=-\left(\frac{1}{k}\right) {\Im m} \tau(E+i0^+). 
\label{intro6}
\end{eqnarray} 

The solution to the Faddeev model is thought to be described in terms 
of a bound state two particle energy of \begin{math}  -E_B  \end{math} 
by the so-called\cite{Berezin:1961,Mead:1991} renormalized expression
\begin{equation}
\tau (z)=\left[\frac{4\pi }{\ln (-E_B/z)}\right] 
({\rm dimensional\ transmutation}).
\label{intro7}
\end{equation}
What is remarkable about Eq.(\ref{intro7}), is that the energy scale 
needed to make the logarithm argument dimensionless cannot be uniquely 
determined by the Hamiltonian in Eq.(\ref{intro1}) since there is no 
combination of \begin{math} \hbar  \end{math} and \begin{math} \mu \end{math}
that has the physical dimensions of an energy. Dimensional transmutation
is thought by some\cite{{Coleman:1973}} to exist in a pure Yang-Mills 
quantum field theory of glue. But it is difficult to construct theories 
that are reasonable that do not obey the usual Abelian multiplicative group 
involved in changing physical units. To conclude that Eq.(\ref{intro7}) has some 
validity one must introduce in some form a length scale which smears out 
the \begin{math}  \delta ({\bf r}) \end{math} in Eq.(\ref{intro1}). 
Without such smearing, the rigorous solution to the Faddeev model is 
\begin{math}  \tau (z)=0 \end{math}, i.e. {\em the two particles neither 
bind nor scatter}. 

In Sec.\ref{gst}, the general theory of scattering amplitudes are reviewed 
with an eye toward describing how such amplitudes vary with energy.
The particular case of a separable interaction is discussed in Sec.\ref{si}. 
For the Faddeev model, it is shown that there is no binding and no scattering 
in Sec.\ref{tfm}. The idea of renormalizing the model to get a bound state 
energy is also discussed. The actual value of the {\em dimensionally transmuted 
energy} is not very well defined., i.e. the renormalized solution is not 
rigorously valid. This point has also been discussed in \cite{Cabo:1998} who
suggest that any apparent anomalous breaking of scale invariance
in the problem is in fact an explicit symmetry breaking due to the 
introduction of a regulator which breaks it. Our exact solution of the
problem confirms this. In the concluding Sec.\ref{conc} the notion of dimensional 
transmutation is further discussed.

\section{General Scattering Theory \label{gst}}

With \begin{math} z  \end{math} as the complex energy in the upper half 
complex plane \begin{math}  {\Im m}  \  z>0  \end{math} ,  the 
scattering matrix for a Hamiltonian decomposition 
\begin{equation} 
{\cal H}=H+V 
\label{gst1}
\end{equation}
is given by 
\begin{eqnarray} 
G(z)=\left[ \frac{1}{z-H} \right] ,
\nonumber \\ 
T(z)=V+VG(z) T(z).
\label{gst2}
\end{eqnarray}
If we fix the scattering amplitude at a reference energy 
\begin{math}  z_0 \end{math}, then the scattering potential 
\begin{equation} 
V=\left[\frac{1}{1+T(z_0)G(z_0)}\right] T(z_0) 
\label{gst3}
\end{equation}
can be eliminated from the scattering Eq.(\ref{gst2}) so that
\begin{equation} 
T(z)=T(z_0)+T(z_0) \left\{  G(z)-G(z_0) \right\} T(z) .
\label{gst4}
\end{equation}
 Eq.(\ref{gst4}) shows in a general manner how it is possible to slide the 
scattering amplitude  from one energy scale 
\begin{math}  z_0  \end{math} to another energy scale 
 \begin{math}  z  \end{math}.  

\section{A Separable Interaction  \label{si}}

A separable two body interaction model producing a possible bound 
state may be taken to be 
\begin{equation} 
V={\cal V}  \left|  v  \right>  \left<  v  \right|  .
\label{si1}
\end{equation}
The scattering amplitude thereby has the form 
\begin{equation} 
T(z)={\cal T} (z) \left|  v  \right>  \left<  v  \right|  .
\label{si2}
\end{equation}
For the separable interaction, Eq.(\ref{gst2}) reads    
\begin{eqnarray} 
g(z) = \left<  v \right| G(z) \left| v \right> , 
\nonumber \\ 
{\cal T} (z) = \left[   \frac{\cal V}{1-{\cal V}  g(z)  }    \right].     
\label{si3}
\end{eqnarray}
Employing 
\begin{equation} 
{\cal G}(z,z_0)=\left<  v \right| G(z) - G(z_0) \left| v \right>  
=g(z) - g(z_0), 
\label{si4}
\end{equation}
the energy scaling of the scattering amplitude is given by  
Eq.(\ref{gst4}) in the separable interaction form
\begin{equation} 
{\cal T}(z)   = \left[  \frac{ 1}{ 1-{\cal T} (z_0)  {\cal G}(z,z_0)  }   \\\right] 
{\cal T}(z_0). 
\label{si5}
\end{equation}
Finally we may define the spectral weight of the separable potential via 
\begin{eqnarray} 
Q(E) =  \left<  v \right|  \delta (E-H)     \left|  v  \right>  , 
\nonumber \\ 
q(t) = \int_0^\infty Q(E) e^{-iEt/\hbar } dE = 
\left<  v \right|  e^{-iHt/\hbar }   \left|  v  \right> ,      
\label{si6}
\end{eqnarray}
so that 
\begin{eqnarray}
g(z)=-\left(\frac{i}{\hbar }\right)
\int_0^\infty e^{izt/ \hbar} q(t) dt,    
\nonumber \\ 
g(z)=\int_0^\infty \left[ \frac{ Q(E)dE } {z-E} \right] ,         
\label{si7}
\end{eqnarray}
and 
\begin{eqnarray} 
{\cal G}(z,z_0) = \int_0^\infty Q(E) \left[ \left( \frac{1}{z-E} \right)  -  
 \left( \frac{1}{z_0-E} \right)     \right] dE ,      
\nonumber \\ 
{\cal G}(z,z_0) = \left(\frac{i}{\hbar }\right)
\int_0^\infty \left[e^{iz_0t/\hbar} - 
e^{izt/\hbar} \right]q(t) dt,\ \ 
\label{si8}
\end{eqnarray}
determines the sliding energy scale.

\section{The Faddeev Model   \label{tfm}  }

In two spatial dimensions, the Faddeev model is described as the 
potential model 
\begin{equation} 
{\cal H} = - \left( \frac{\hbar ^2}{2\mu } \right) 
\left[  \Delta +\epsilon \delta ({\bf r}) \right] = 
H+{\cal V}\delta ({\bf r}) ,   
\label{tfm1}
\end{equation}
wherein \begin{math} \epsilon  \end{math} is a dimensionless 
coupling strength. The short ranged potential is thereby 
\begin{eqnarray}
V({\bf r}) = {\cal V}  \delta ({\bf r}) = 
- \left( \frac{\hbar ^2}{2\mu } \right)   \epsilon \delta ({\bf r}), 
\nonumber \\ 
V({\bf r})\psi ({\bf r}) = {\cal V}  \delta ({\bf r}) 
\psi ({\bf 0}) = 
\int \left<{\bf r}\right|V\left| {\bf r}^\prime \right>  
\psi ({\bf r}^\prime )   d^2 {\bf r} ^\prime ,  
\nonumber \\ 
\left<{\bf r}\right|V\left|{\bf r}^\prime \right> = 
{\cal V} \delta ({\bf r})\delta ({\bf r}^\prime ) = 
{\cal V}\left<{\bf r}|v\right>\left< v |{\bf r}^\prime \right>, 
\nonumber \\ 
\left<{\bf r}|v\right> =\delta ({\bf r}) 
= \int \left<{\bf r}|{\bf k}\right> \left<{\bf k}|v\right> 
\left[\frac{d^2{\bf k}}{(2\pi )^2}\right], 
\nonumber \\ 
\left<{\bf r}|{\bf k}\right> = e^{i{\bf k\cdot r}} 
\ \ \ \Rightarrow \ \ \ \left<{\bf k}|v\right>=1.  
\label{tfm2}
\end{eqnarray}
One may now compute the rigorously exact solution for this Faddeev 
problem. 

\subsection{Sliding Energy Scale  \label{ses}} 

Employing Eqs.(\ref{si6}) and (\ref{tfm1}) yields the decay amplitude 
\begin{equation}
q(t)=\int \exp \left(-\frac{i\hbar k^2 t}{2\mu }\right) 
\left| \left<{\bf k} | v \right> \right|^2 
\left[\frac{d^2{\bf k}}{(2\pi )^2}\right].   
\label{ses1}
\end{equation} 
From Eqs.(\ref{tfm2}) and (\ref{ses1})  
\begin{equation}
q(t)=\left(\frac{\mu}{2\pi i t \hbar}\right) .   
\label{ses2}
\end{equation} 
In virtue of Eqs.(\ref{si8}) and (\ref{ses2}) one computes 
\begin{eqnarray} 
{\cal G}(z,z_0) = \left(\frac{\mu }{2\pi \hbar^2}\right)
\int_0^\infty \left[e^{iz_0t/\hbar} - 
e^{izt/\hbar} \right] \frac{dt}{t}\ ,  
\nonumber \\  
{\cal G}(z,z_0) = \left(\frac{\mu }{2\pi \hbar^2}\right) 
\ln \left[\frac{z}{z_0}\right].  
\label{ses3}
\end{eqnarray}
Let 
\begin{equation}
\tau(z)=\left(\frac{2\mu }{\hbar^2}\right) 
{\cal T}(z). 
\label{ses4}
\end{equation}
Eqs.(\ref{si5}), (\ref{ses3}) and (\ref{ses4}) now read 
\begin{equation}
\tau(z)=\left[\frac{\tau(z_0)}{1-\big[\tau(z_0)/4\pi \big]\ln(z/z_0)} \right]. 
\label{ses5}
\end{equation}
Eq.(\ref{ses5}) is central to the solution of this Faddeev model. 
\begin{equation}
\frac{1}{\tau(z)} = \frac{1}{\tau(z_0)} - 
\left[\frac{1}{4\pi }\right]\ln \left(\frac{z}{z_0}\right). 
\label{ses6}
\end{equation}
Having chosen the energy scale \begin{math} z_0 \end{math}, the 
scattering amplitude at any energy \begin{math} z \end{math} may 
be found from its value at that chosen scale. There is no guarantee, however,
that that value, at   \begin{math} z_0 \end{math} is nonzero. The characteristic logarithm appears in the denominator 
of Eq.(\ref{ses5}).

\subsection{Fixed Energy \label{fe}}

To evaluate the scattering amplitude at an initial complex energy 
\begin{math} z_0 \end{math}, one must evaluate Eqs.(\ref{si3}), (\ref{tfm2}), 
(\ref{ses1}) and (\ref{ses4}) to arrive at 
\begin{eqnarray}
\tau(z_0) = -
\left[ \frac{\epsilon}{1+(\hbar^2/2\mu)\epsilon g(z_0)} \right] 
=  -\left[ \frac{\epsilon}{1+\epsilon I(z_0)} \right] , 
\nonumber \\ 
I(z_0)=-i\left(\frac{\hbar^2}{2\mu}\right)
\int_0^\infty q(t)e^{iz_0t/ \hbar } dt.
\label{fe1}
\end{eqnarray}
From Eqs.(\ref{ses2}) and (\ref{fe1}), one has for the Faddeev model 
\begin{eqnarray}
I(z_0)=-\left[\frac{1}{4\pi} \right] 
\int_0^\infty e^{iz_0t/\hbar}  \left(\frac{dt}{t}\right)
\nonumber \\ 
|I(z_0)|=\infty \ \ \ {\rm for} \ \ \ {\Im m}\ z_0>0  
\label{fe2}
\end{eqnarray}
due to the logarithmic divergence in the time integral  as 
\begin{math} t\to 0  \end{math}. Thus, we have 
\begin{math}  \tau(z_0)=0   \end{math} and then in virtue of Eq.(\ref{ses5}) 
the following\cite{Friedman:1972}: 
\par \noindent 
{\bf Theorem:} {\em The Faddeev singular potential does not scatter  nor 
does it bind the particles, i.e.}
\begin{equation}
\tau(z)=0 \ \ \ {\rm for} \ \ \ {\Im m}\ z>0.
\label{fe3}
\end{equation}
Since the divergence in \begin{math} I(z_0)  \end{math} is merely 
logarithmic, one might seek to {\em evade this rigorous theorem} by means 
of an intuitive {\em renormalization viewpoint}.

\subsection{Theorem Evasion \label{te}}

In order to evade a rigorous theorem, it is required to relax standards of 
mathematics. A formal derivative of the \begin{math} I(z_0) \end{math} 
function in Eq.(\ref{fe2}) yields 
\begin{equation}
I^\prime (z_0)=-\left[\frac{i}{4\pi \hbar} \right]
\int_0^\infty e^{iz_0t/\hbar}dt = \left[\frac{1}{4\pi z_0} \right].
\label{te1}
\end{equation}
The general solution of Eq.(\ref{te1}) can be expressed in terms of a 
{\em large energy cut-off} \begin{math} \Lambda  \end{math} as 
\begin{equation}
I(z_0,\Lambda )=\left(\frac{1}{4\pi }\right)
\ln \left[\frac{-z_0}{\Lambda }\right].
\label{te2}
\end{equation}
Note that \begin{math}  |I(z_0,\Lambda )| \to \infty  \end{math} as 
\begin{math} \Lambda \to \infty  \end{math} in agreement with 
Eq.(\ref{fe2}). Now the scattering amplitude with a high energy cut-off 
follows from Eqs,(\ref{ses6}) and (\ref{fe1}). It is   
\begin{equation}
\frac{1}{\tau(z)}=-\left(\frac{1}{\epsilon}\right)-
\left(\frac{1}{4\pi }\right)\ln \left[\frac{-z_0}{\Lambda }\right] 
-\left(\frac{1}{4\pi }\right)\ln \left[\frac{z}{z_0}\right]. 
\label{te3}
\end{equation}
Now we may consider the so-called renormalization prescription: 
\par \noindent 
(i) Choose a bound state energy \begin{math} z_0= - E_B \end{math} 
so that the sum of the first two terms on the right hand side of 
Rq.(\ref{te3}) vanish; i.e. 
\begin{equation} 
-z_0=\Lambda e^{-4\pi/\epsilon }\equiv E_B .  
\label{te4}
\end{equation} 
\par \noindent 
(ii) The scattering amplitude has a singularity at the bound 
state energy 
\begin{equation} 
\tau (z) = \left[\frac{4\pi}{\ln(-E_B/z)}\right].  
\label{te5}
\end{equation} 
\par \noindent 
(iii) The bound state energy appearing in the denominator logarithm 
of Eq.(\ref{te5}) has appeared from nowhere since there is nothing 
with dimensions of energy in the Faddeev model. This paradox is known 
as {\em dimensional transmutation}. However the bound state energy 
can be whatever one wishes it to be. The renormalization 
limiting process, 
\begin{equation} 
E_B=\left[\lim_{\Lambda \to \infty\ {\rm and}\ \epsilon \to 0^+}\right]   
\left[\Lambda \exp \left(-\frac{4\pi}{\epsilon}\right)\right],
\label{te6}
\end{equation} 
\par \noindent 
gives one no idea of the actual value of the bound state energy 
\begin{math} E_B \end{math}. 

The paradox is resolved when it is realized that for the rigorous solution 
of the Faddeev model in Sec.\ref{fe} {\em there is no bound state energy 
and there is no scattering}. A mathematical theorem is very difficult to evade.

Even in very recent literature this seems to have remained a point
of confusion. Padmanabhan in his otherwise delightful book\cite{Padmanabhan:2015}
of 2015, writing in his discussion of this problem in Chapter 10, describes the problem
as ``ill-defined'', but then ``taking a clue from what is done in quantum field
theory'' goes on to try to find the scattering cross section for an attractive
delta function potential in 2 dimensions anyway. This is eventually
expressed formally in terms of the ``bound state energy''. The determination of
this putative bound state energy requires an
additional process outside the mathematics of the problem where ``one performs
an experiment to measure some observable quantity (like the binding energy) of
the system as well as some of the parameters describing
the system (like the coupling constant)''. Of course this project is not going
to work if there is no bound state at all, so his formal solution is in fact unphysical,
in accord with his initial intuition.
The theory is as incapable of delivering a cross section as it is of delivering a bound
state energy -- and for the same reason: any quantity with dimensions of length or
energy must be found by extending the model to some larger theory which can provide
quantities with the needed dimensions. The one given simply does not suffice. Similar arguments
are given throughout the literature\cite{Mead:1991,Gosdzinksy:1991,Holstein:1993,Hans:1983}, though rarely
with Padmanabhan's clarity in showing how an unphysical result can arise.

\section{Conclusion \label{conc}}

The Faddeev two body bound state model has been discussed as an example of a 
QCD inspired model thought by some to exhibit {\em dimensional transmutation}. 
This simple model was solved exactly and the growth of a specified dimensional 
energy scale is shown to be something of an illusion.  

The argument concerning quantities with physical dimensions again occurs 
in the pure Yang-Mills quantum field theory of pure glue. It does not seem 
possible to generate physical quantities with physical dimensions if such 
dimensions do not occur in the microscopic Hamiltonian. The Faddeev model is 
a particularly simple model that can be employed to discuss the general problem.

\section*{Acknowledgments}

J. S. would like to thank the United States National Science Foundation for support under PHY-1205845.

\vfill \eject

\end{document}